# A.I. In All The Wrong Places

Marc Böhlen[1], Ruolin Chen[1], Xiaoxu Dong[1], Srikar Gopaladinne[1], Hemanth Gorla[1], Divya Kandukuri[1], Sean Mansfield[1]

[1] University at Buffalo, USA

`marcbohlen@protonmail.com`

**Abstract.** This text describes experiences gained across a two-year test period during which two generations of Generative Artificial Intelligence (A.I.) systems were incorporated into an interdisciplinary, university level course on A.I. for art and design practices. The text uses the results from the courses to reflect on new opportunities for generative systems in art and design while considering traps and limits.

## 1 Artificial Intelligence Explosion

In a 2017 blog post, the CEO of NVIDIA, the firm responsible for the design of much of the hardware that generative A.I. systems operate on, claimed that the combination of processing power, cloud access and A.I. algorithms learning from raw data, are fueling a *Cambrian explosion* [1] of A.I. Technology. Being in the midst of a Cambrian explosion, even if only a metaphor, is intoxicating and disorienting at once. In Earth Science, the Cambrian explosion describes a period of sudden increase of complex life forms some 550 Ma. However, the presumable explosion raises questions, including the role of fossilization potential [2], suggesting that the assessment of the explosion might be impacted by the well-preserved fossil record itself. At any rate, the oldest animal fossils, such as trilobites, predate the Cambrian period, reminding us that even explosive events have histories and contexts.

Earth scientists have the advantage of hindsight. There is no historical distance that allows us to definitively assess whether the current times are Cambrian explosions of A.I. technology. The goal of this paper is to reflect with some nuance on current generative A.I. systems impact in art and design education, an admittedly minor subset of AI domains, and to suggest where these novel systems might be differently generative, and where it might make more sense to forsake them.

## 2 Early Generative Design Experiments

Artists and designers have long been attracted to the potential of generative systems. The concept of *Cellular Automata*, first devised by Stansilaw Ulam and John von

Neumann [3] received over time intense interest in art, design and architecture as a method of generating change within limits. Cellular automata typically consist of a regular grid of cells, where each cell is in a distinct state. New states are generated based on a set of local transition rules, typically mathematical equations, and the layout of cells adjacent to a given cell. The lure of cellular automata stems from the fact that complex nonlinear behavior can emerge from a simple set of rules, allowing for surprising forms of change over time, mimicking the dynamics of natural processes, or, with some artistic license, life itself. The *Game of Life*, invented by John Conway [4], was designed as a cellular automaton. An informal version of cellular automata logic within experimental architectural design can be found in Cedric Price's *Generator Project* [5]. While this environment was never realized, Price hoped to use algorithmically enabled flexible arrangements of cube-like elements to create temporary dwelling structures.

While cellular automata can produce surprising behaviors, they are limited by the lattice arrangement and governing neighborhood rules. *Genetic algorithms* are more flexible generative systems. While the origins of genetic algorithms reach back to Turing's concept of learning machines, they were popularized by John Holland in the 1970s [6]. Genetic algorithms use optimization procedures inspired by biological systems to create, across multiple iterations, novel features through a combination of mutations and crossover operations, the outcome of which are evaluated by a fitness function [7]. Inspired by a Darwinian view of the world, the fitness function defines the criteria that every offspring of the generative process is assessed by. The adaptive savvy of genetic algorithms was put on display in the design of an antenna for NASA's ST5 spacecraft. By specifying the design constraints in terms of a fitness function, the engineers were able to craft a genetic algorithm that evolved radically novel antenna designs with, surprisingly, better performance properties than conventional spacecraft antenna [8].

Karl Sims was perhaps one of the first artists to demonstrate the uncanny ability of genetic algorithms to create graphical objects capable of emergent, complex behaviors [9]. Sims' *Virtual Creatures* research devised a set of fitness functions and physical constraints to evolve simple geometric forms into complex assemblages of interconnected parts that behaved like goal-driven living systems. These creatures were capable of swimming, walking, jumping and competitive play in their simulated environment, and performed behaviors that were recognizable as live-like and yet eerily distinct from Earth-specific life forms.

*Neural networks* are computer based approximations of biological networks. They can adapt their outputs based on exposure to information they are exposed to and learn to emulate patterns detected in that information. Neural networks have largely replaced genetic algorithms as a site for artistic experimentation in generative systems, and in some cases, the internal processes themselves become sites of inquiry, as Audry describes in a survey of the field [10]. The list of artists using generative A.I. is long and expanding, and the results not always interesting. An early example of the tensions that can emerge in works that deploy generative A.I. to emulate existing art genra such as painting, comes to the foreground in *Edmond de Belamy*[1], created by a generative adversarial network by the arts-collective *Obvious* in 2018, and later auctioned off by Christie's. The fact that image generators produce visual artifacts that are difficult to distinguish from human visual production became a prominent talking point after the winner of the 2023 Sony World Photography competition revealed that his moody black and white 'photograph' of two women was in fact the product of an A.I. art generator. The artist, Boris Eldagsen, refused the award, stating that "A.I. is not photography"[2]. A less combative response to the pressures produced by the increasing imitative capacity of generative systems can be found in Pablo Delcan's project *non-AI art generator* [11]. Here, the artist takes on the role of the obedient image generator. Responding to prompts from visitors to his website, he deftly created what was requested, in simple line drawings, dashing thick black lines on a white background. While the project began in gist, it became seriously clear that the approach could not scale, as hundreds of prompts sat idle awaiting a response. Imitating machinic behavior is a challenge. Certainly over the long term.

## 3   Large Generative A.I. Models

Generative A.I. is very much in flux. New methods and techniques are redefining limitations experienced in systems presented only a few years ago. The contribution of transformer networks that capture context much better than earlier neural network systems, is at the core of this change. The next section will focus on transformer based language generation for interactive chatbots and the DALL-E series image generators we based the course experiments on.

All language models used in our courses are versions of the GPT (generative pre-trained transformers) family released by OpenAI. At the core of all of these models are a collection of transformer neural networks based on the attention mechanism. As opposed to earlier language generation systems based on recurrent neural networks [12], transformer networks have neither recurrence nor convolution steps [13]. Instead, transformer networks operate with encoder, decoder and attention processes. Encoders convert text into vector representations while taking word position into account. Decoders use those encoded vector representations to predict the next word based on previous words with the help of attention. The attention mechanism is a mathematical operation that allows the network to take varied and longer sequences of tokens into account when it makes a prediction.

Paying attention to longer input sequences in the generation of predictions allows for context and linguistic nuance to propagate through the network. Transformer networks are large systems with billions of parameters, and they require training operations on massive datasets. GPT models outperform previous natural language generation models by wide margins and are at the core of chatbots capable of cogent content summarization, language translation and text generation. OpenAI's GPT series differ in model size and complexity, and the resultant ability to perform text synthesis.

---

[1] https://www.theverge.com/2018/10/23/18013190/ai-art-portrait-auction-christies-belamy-obvious-robbie-barrat-gans, last accessed 2023/01/10.
[2] https://www.bbc.com/news/entertainment-arts-65296763, last accessed 2023/01/10.

GPT-2, released in 2019, contained 774 million parameters [14] and was able to produce believable text and plausible propaganda [15], according to OpenAI research partners. GPT-2 was a rather coarse text generator that required substantial fine-tuning and was trained on data acknowledged to be biased in embarrassing ways [16]. GPT-3, released in 2020, contains some 175 billion internal parameters in the davinci version and requires no fine-tuning, allowing for direct, zero-shot text production. GPT-3 is capable of modest arithmetic operations and is able to produce news articles that human evaluators at the time struggled to identify as machine produced [17]. GPT-4, released in 2023, is assumed to be vastly larger than its predecessor with perhaps an order of magnitude more internal parameters, though the details of this model have not been made public [18]. GPT-4 is a multi-modal model that can process image and text inputs and produce text outputs [19]. As of 2024, it is the most sophisticated of the GPT series and capable of passing a simulated bar exam [20] that includes "a standardized 200-item test covering six areas (Constitutional Law, Contracts, Criminal Law, Evidence, Real Property, and Torts)"[21], suggesting substantial craftiness. Regardless, GPT-4 is the first of the GPT models to be subject to alignment adjustments, intended to sync the model with human-like preferences, rendering the outputs "more truthful and less toxic" [22]. Additionally, GPT-4 is more responsive to input prompting, making it more easily steerable by users and adaptable to specific use case scenarios, even without the fine-tuning required in earlier GPT versions.

The DALL-E image generators take text prompts as input and produce images corresponding to those text prompts. The DALL-E series is differentiated along procedures that increasingly improve the transfer of the idea expressed as text into an image of increasing quality and resolution. The transformer based encoding and decoding scheme together with the creation of a joint latent space that can transfer the content of language, within limits, onto an image that can express a similar idea, is perhaps the most interesting part of the operation.

For example, DALL-E-2 has a training process that creates a joint representation space for texts and images. This training process makes use of a large proprietary and secretive labeled training set, *WebImageText*. Using that labeled training set, a process called *diffusion* [23] transforms a text prompt into a high-resolution image. Diffusion starts with a random noise image, and refines that image over many steps until it resembles what the text prompt describes. That iterative process is guided by a neural network, U-net, developed for image segmentation in medical images. If you tell a diffusion network to create an astronaut on a horse, it will create a photorealistic image of a generic astronaut on a horse.

DALL-E-3 is structured similarly to DALL-E-2. However, it includes and an additional model, image captioner, to expand on the details of an input caption from the image-text training set. For example, the original caption "a jar of rhubarb liqueur sitting on a pebble background" becomes "rhubarb pieces in a glass jar, waiting to be pickled. the colors of the rhubarb range from bright red to pale green, creating a beautiful contrast. the jar is sitting on a gravel background, giving a rustic feel to the image". With this augmented input, the image decoder can produce more realistic and differentiated images, outperforming other text-to-image models by wide margins [24]. Interestingly, hobbyist artists have taken issue with some of the features of the improved DALL-E-3 model, suggesting that it lacked the 'style' of its predecessor [25], [26], looking for a human specific trait in all the wrong places.

Prompting techniques applied to GPT-4 and DALL-E-3 level generators is crucial to the quality of the produced responses. However, while the significance of prompt design is widely recognized, the practical crafting of prompts remains a field of experimentation. Mixing vocabulary with artistic license, a writer, not an engineer, defined the term now used for crafting prompts, to wit, *prompt engineering* [27]. Part of the prompt engineering process requires one to iteratively craft prompts, evaluating the output at each step. That human-machine discourse is key to improving the responses of text-to-image generators. Human computer interaction designers identify this process as a *creativity support tool* [28], bypassing the rather significant question of where the creativity in fact resides in the process. Because the linguistic quality of prompts manifestly impacts the quality of the result [29], users with honed language skills can coax synthetic text and image generators to produce differentiated results, creating, perhaps unexpectedly, a novel niche for sophisticated writing in the age of generative A.I.

## 4     Generative A.I. in an Art & Design course

The course *Critical Machine Learning* [30] is designed to immerse students of Art, Design, Architecture and Computer Science into the logics of machine learning. The course has two parts. The first one surveys early machine learning classic algorithms, such as Naive Bayes, Random Forests, Support Vector Machine and simple Neural Networks. The second part of the course switches to elements of large language models, Encoder, Decoders and Attention mechanisms. While the concepts are discussed in detail, the practical code work we perform largely works with existing math and machine learning libraries, including Scikit, Keras and Pytorch on bespoke virtual machines in the cloud. We interface to OpenAI's text and image generation engines with APIs and post-process the results to suit our needs. Given the diverse student skill levels and backgrounds, we typically work in groups, specifically in the later part of the course. The focus across both course modules is twofold. First to grasp the basic ideas behind machine learning algorithms and second, to stress test them through playful inquiry, including subjecting them to experimental datasets such as the first machine learning compatible collection of tropical plants [31].

### 4.1     Course Artifacts

The first version of this course in 2022 used GPT-2 and Dall-E to explore generative A.I. systems. The goal of the course's group assignment was to create a graphic novel using GPT-2 for text and Dall-E for image generation. Graphic novels are distinct from comic books and cartoons in that they are perhaps closer to literary production, though the comics' community might disagree with that assessment. The example graphic novel shared with students as inspiration was Art Spiegelman's *Maus* [32], a graphic novel about the Holocaust, with Jews rendered as mice and Nazis as cats. Students experimented with GPT-2 and attempted to query its capacity to tell a story based on input prompts organized in jupyter notebooks on *Colab*[3], with disappointing results. After initial experiments, they attempted to use GPT-2 more in a prompt-response mode. They generated a story from one seed prompt and then responded in turn to those outputs, letting the design of the story plot emerge from that interaction. The emergent narrative lacked a clear thread, yet created a semblance of discourse across two or three interactions.

Dall-E was to deliver images that corresponded to the individual story plot elements. However, the output of Dall-E in 2022 was deemed inadequate for even modest visual ambitions and discarded out of hand as a viable image creation option. Since the imagery produced by Dall-E was unusable, students replaced the machinic images with their own hand-drawn imagery and integrated GPT-2 generated text blocks into those drawings, adjusting the visual graphic style to the textual flavor of the text. An example of the final collaboration between student and machine is given in Fig. 1.

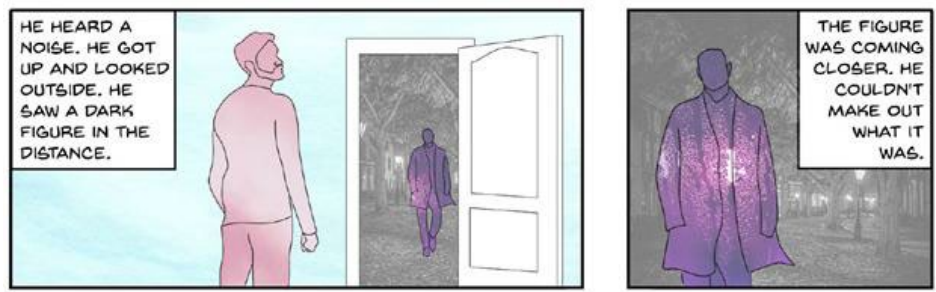

Fig. 1 John Fishetti in collaboration with GPT-2, 2022.

By 2023, when the course was offered a second time, OpenAI had already updated the GPT and Dall-E ecology and made GPT-3 and Dall-E2 publicly available. This time, the theme of the group project was left up to the students, with one boundary condition. They were to implement the project as a public website. This requirement was intended as an opportunity to push students to reflect on generative A.I. in the public realm, as an object that can operate outside their control. Where A.I. is consumed matters materially for the reach of A.I. Both gains and harms are amplified by accessibility and distribution, and this insight is better gained in a sandbox than in the real world.

In order to support this experimentation space, we created custom virtual machines and used *Streamlit*[4] to build a python-supported frontend, eliminating the necessity to employ distinct languages for frontend and backend. The backend, consisting of python scripts, facilitated the integration of machine learning libraries and the retrieval and organization of detailed prompts provided by the frontend. These organized prompts are subsequently transmitted to GPT-3 and DALL-E2 through API calls. The returned results were then integrated into the website dynamically, allowing for an interactive experience of generative A.I. in action.

---

[3] https://colab.research.google.com/, last accessed 2024/26/01.
[4] https://streamlit.io/, last accessed 2023/01/10.

Our design and ideation process included long discussions on the types of experience we should consider. The lack of critical and informed texts on the subject of transformer-enabled generative A.I. systems in 2022 and early 2023 necessitated a more impromptu literature survey for context materials outside of the developer and engineering communities. We surveyed works on the promises and perils of A.I., including essays on how A.I. might impact the workplace [33], speculative essays pondering what kind of 'mind' ChatGPT has [34] and the potential for propaganda through chatbots [35].

Armed with a bespoke development environment, a basic understanding of generative A.I. and the contexts in which it operates, the students embarked on a group project to probe some of the potentials of generative A.I. After lengthy discussions and several experiments, they settled on a cook book, more specifically a A.I. enabled recipe generator, the *Neural Recipe Box*. Fig. 2 shows a screen shot of the website's landing page.

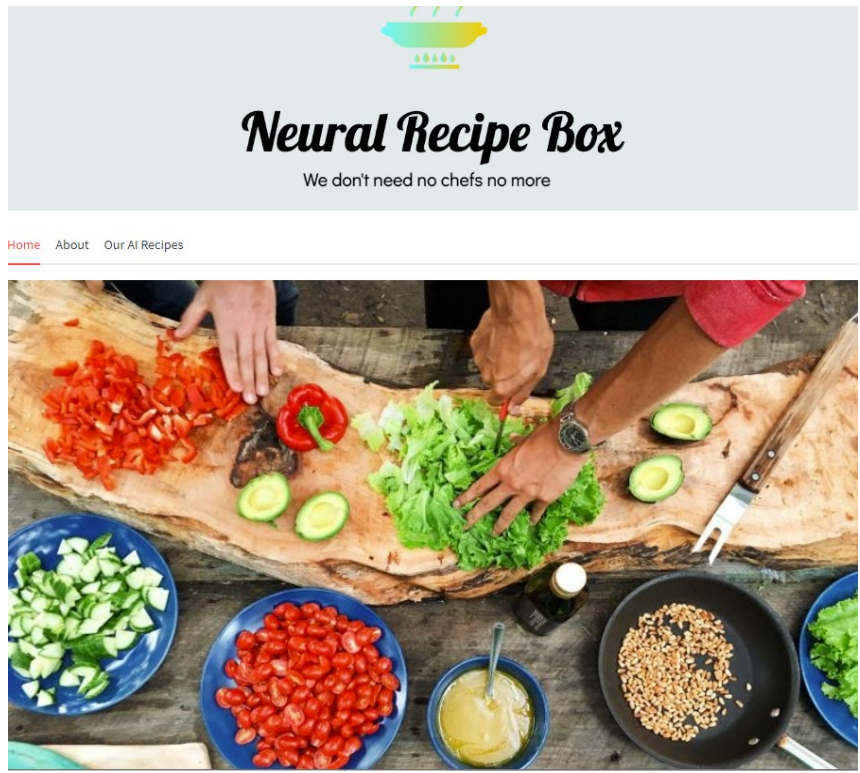

Fig. 2. Landing page of the Neural Recipe Box.

The recipe generator was embedded in a website where visitors could select a mood, culinary preferences, dietary restriction, cooking conditions, etc., and then let the A.I. generator produce from those inputs a proposed recipe, including cooking instructions, calories and an account on the ingredients used in the preparation of the dish. From a prompt engineering perspective, the crafting of the inputs from the web selected fields was challenging. Here is the detailed eight point requirement list included in the prompt sent via API call to the GPT engine.

Table 1. prompt engineering ingredients for the Neural Recipe Box.

| |
|---|
| 1 - Generate a recipe that incorporates the following factors: ingredients - tofu, dish type - snacks, cuisine type - Indian cuisine, food-restrictions - vegetarian, meal-type - Lunch, mood - Excited, texture of the food - grainy, smell - Fermented, color - Turquoise, temperature - , heating method - Frying, and calories - 100-200 calories. |
| 2 - Check if the recipe is edible and safe and assure the reader about its safety with a heading. |
| 3 - Make the recipe compelling with cooking instructions like the blog of a famous food blogger which targets an advanced food connoisseur to push the boundaries of culinary savoir-faire. |
| 4 - Add a paragraph about the place of origin of the dish and make sure the recipe focusses more on the usage of tofu and translate them in the local language of place of origin and explain their cultural significance and any obscure facts about them. |
| 5 - Create a table showing the macronutrient and micronutrient composition of the recipe based on the ingredients and their quantities used. |
| 6 - Add a paragraph on the nutritional value and medical benefits this recipe provides by the research studies in the style of health information site like "Healthline". |
| 7 – Add three comments from blog users with realistic names on the above recipe with positive, negative, and mixed sentiments. |
| 8 - Make sure to use heading names exactly as listed: Recipe Name, Ingredients, Dish Type, Cuisine Type, Food Restrictions, Meal Type, Mood, Texture, Smell, Color, Temperature, Heating Method, Calories, Instructions, Safety Assurance, Origin and Cultural Significance, Nutritional Composition, Nutritional Value and Medical Benefits. |

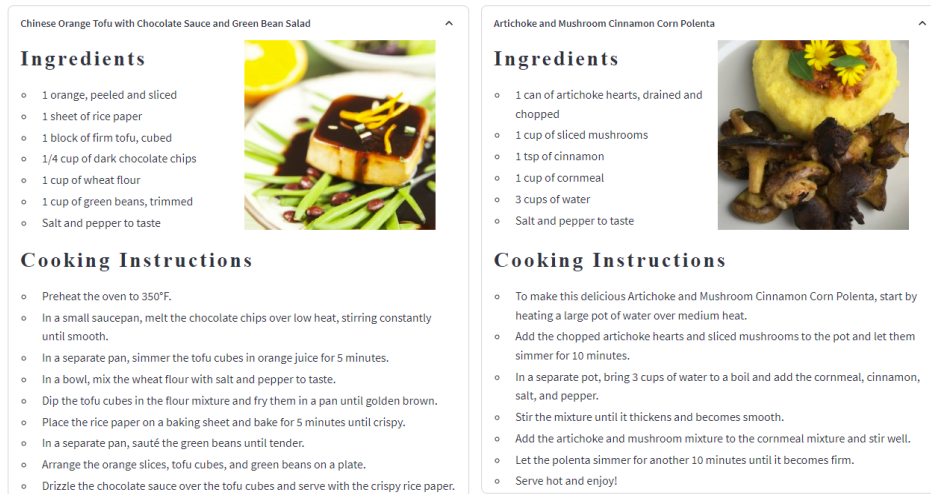

Fig. 3 Two recipes from the Neural Recipe Box collection.

We created a small collection of meals with the Neural Recipe Box, exploring the affordances of different GPT and DALL-E versions. Fig. 3 shows two of the results from this collection. Table 2 shows an overview of results using GPT-3.5 versus GPT-4, and Table 3 compares results from DALL-E-2 versus DALL-E-3.

Table 2. Titles and cooking instructions from GPT-3.5 and GPT-4.

|   | GPT-3.5 title | GPT-4 title | GPT-3.5 instructions | GPT-4 instructions |
|---|---|---|---|---|
| 1 | Thai Chicken Purple Curry | Thai Purple Sticky Chicken Seafood Platter | 1. Heat a pan over medium heat and add the garlic and onion. Sauté until fragrant.<br>2. Add the chicken breast and cook until browned on all sides.<br>3. Add the curry paste, fish sauce, and coconut milk. Stir until well combined.<br>4. Cover the pan and let it simmer for 20-30 minutes or until the chicken is fully cooked.<br>5. Add the bell peppers and Thai basil leaves. Cook for another 5 minutes until the vegetables are tender.<br>6. Season with salt to taste.<br>7. Serve with steamed rice or noodles. | To push beyond boundaries of culinary savoir-faire, start by sautéing minced garlic and onion in the vegetable oil in a slow cooker. Add sliced chicken breasts and seafood, drizzle gluten-free soy sauce and stir until lightly browned. Add diced purple sweet potatoes, taro, Thai basil, tutti fruity essence, and coconut milk. Keep it slow-cooking until the mixture attains a gooey texture. Just before serving, sprinkle chilled pineapple juice to give a cool 15°C - 20°C temperature. Your Thai Purple Sticky Chicken Seafood platter is ready to elevate the color palette and brunch scene. |
| 2 | Greek Coriander Flaky Stew | Greek Light Coriander Stew | Heat a large pot over medium heat and add onion, stirring until softened. Add garlic, coriander, turmeric, paprika, cumin, and thyme, and cook until fragrant. Add tomatoes, chickpeas, and vegetable broth, and bring to a boil. Reduce heat to low and simmer for 30 minutes, stirring occasionally. Add lemon juice, fresh coriander, and simmer for an additional 10 minutes. Serve hot. | Begin by sauteing the diced white onions in a non-stick pan, followed by minced garlic. Add umami powder for that Extra savory Twist. Mix in chopped tomatoes and slow cook till it acquires a uniform texture. Finally, mix in Greek yogurt and the star of the dish - chopped coriander. Stir well and freeze for a rich, flaky texture. Enjoy every bite of this frozen flaky delight. |
| 3 | French Lentil Soup | La Soupe Bleue de Lentilles | 1. In a large saucepan, heat the olive oil over medium heat.<br>2. Sauté the onions, celery, and carrots until the onions become translucent, about 5 minutes.<br>3. Add the garlic and sauté for another minute.<br>4. Add the chicken broth, bay leaves, and thyme to the saucepan and bring the mixture to a boil.<br>5. Once boiling, add the lentils and let the soup simmer until the lentils are tender, about 30 minutes.<br>6. Remove the bay leaves and season the soup with salt and pepper to taste.<br>7. Preheat the oven to 375°F.<br>8. Transfer the soup to a roasting pan and roast for 5-10 minutes until the top is slightly browned. | For the adventurous and spirited food connoisseur, delve into the gradations of French culinary finesse. Roast the lentils over a medium flame, stirring continuously until crispy. Sauté the onions and garlic in vegetable oil till golden brown, emanating their delightful umami aroma of freshly fried comfort food. Subsequently, this mixture is to be added to the stock with the well-roasted lentils and allowed to simmer till the lentils are wickedly soft. Finally, infuse your creation with a hint of adventurous excitement by incorporating gluten-free blue food colouring. Serve this riveting potage freezing to enjoy its exquisiteness. |

Table 3. Images from DALL-E-2 and DALL-E-3.

| | DALL-E-2 image | DALL-E-3 image |
|---|---|---|
| 1 | 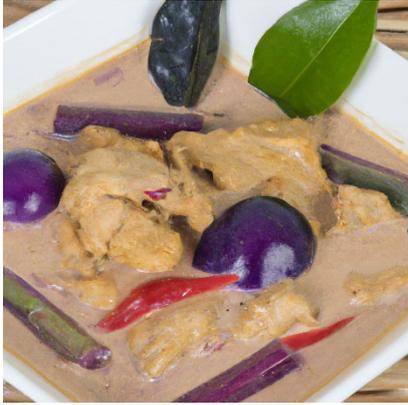 | 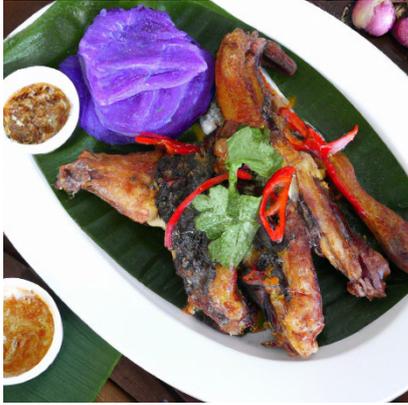 |
| 2 | 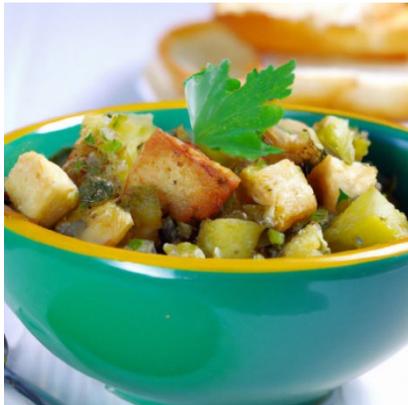 | 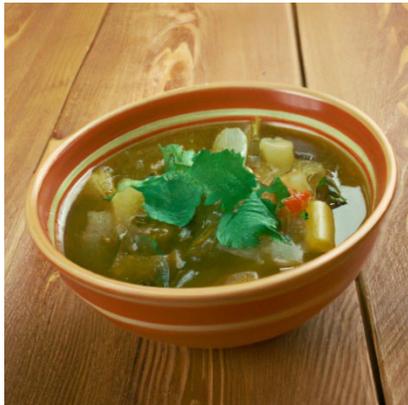 |
| 3 | 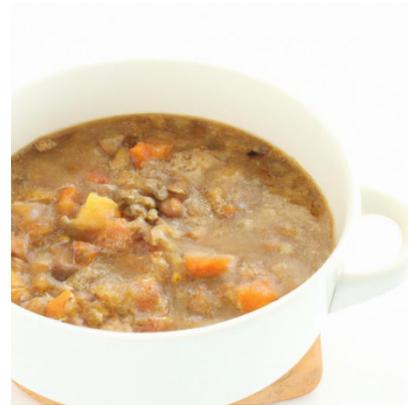 | 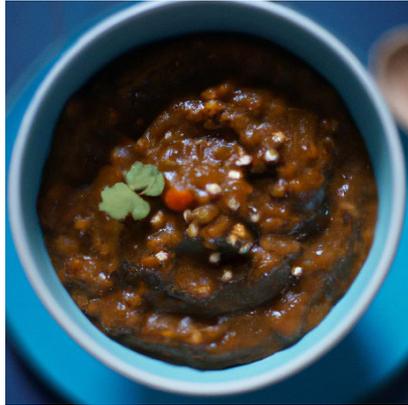 |

Overall, it appeared that GPT-4 followed the instructions more carefully than GPT-3.5. It also created more elaborate recipes for fancier tastes than its GPT-3.5 counterpart. GPT-3.5 prose style was also overall more pedestrian than GPT-4 in this recipe generation experiment. While the newer GPT version delivers more refined recipes and contextual information, the shift to the newer image generator delivers in this use case much less clear advantages. In most cases, we cannot find qualitative differences between the two models. In some cases DALL-E-3 appears to create more detailed backgrounds, finer texture and more nuanced lighting than its predecessor, but these improvements do not always make the recipes more appetizing.

We found the preparation instructions to be inconsistent at times. GPT-3.5 generated preparation instructions seemed more concise than the GPT-4 counterparts. GPT-3.5 instructions were almost always enumerated and presented as short statements while the GPT-4 counterparts were formulated as long running sentences, flourished with flowery verbiage, rendering those results less useful in practical food preparation contexts. Yet who is to blame? Our instructions required the bot to try to write for connoisseurs.

## 5    Discussion

### 5.1    Surprises

Across all our experiments we were surprised by the ability of the text and image generators to concoct suggestions for meals while addressing all the requirements listed in the elaborate prompt input. Moreover, the system was able to continue its inventiveness ad infinitum, devising new recipes at each and every call. As such, it demonstrated a non-human ability to innovate. Addressing all the input requests is a performance a human chef would hardly be able to match, and perhaps refuse to match. The culinary innovations at times crossed boundaries, at least for food traditionists. A proposed Citrusy Vietnamese Pizza might not fare well with pizza purists. At the same time, many proposed recipes felt somewhat derivative, at least in the imagery, of standard cookbook products. Additionally, no inventiveness in presenting the meals themselves was discernable. Perhaps a uniformity of training materials sourced from cookbook photography impacts this outcome.

The surprising abilities we observe confirm what other researchers have observed in their tests of GPT, and in particular the GPT-4 level text generators [36], [37]. It appears that the scale in parameter size, dataset size and training cycles of the newest models is at the core of the observed performance increase. Generative models have been observed by others to exhibit a paradoxical combination of unpredictable capabilities together with high level predictability [37], and perhaps our experiments make more sense seen in this light. Some of the proposed recipes seemed not only quirky but unappetizing. GPT does not seem to gauge how an innovative recipe would taste. That feedback loop is not established, and generative prowess alone cannot answer that question.

We felt unease when reading the dutifully produced comments, arranged as requested into positive and less positive assessments. The system invented names associated with the comments, suggesting an unsettling opportunity to fuel fake reviews. Future iterations of the GPT series, perhaps material for alignment research, might be more keenly attuned to such conditions and programmatically prevent generating this class of misinformation, even when kindly prompted to do so. The appendix includes a complete example of the textual artifacts produced by the Neural Recipe Box, including 'comments'.

Finally, we ventured into our own kitchens and tested several of the recipes concocted by the Neural Recipe Box. We found the flowery GPT-4 recipes, that mostly did not include cooking times, less useful in the kitchen. The functional language dominant in the instructions of GPT3.5 instructions were easy to follow. However, some of the details were misleading. For example, while preparing *Crispy Garlic Roasted Eggplant*, the proposed 30 minute baking period rendered the eggplant not golden brown but burnt black, more reminiscent of a BBQ eggplant preparation. Other inconsistencies were observed during the preparation of GPT-3.5's *French Lentil Soup.* The recipe stipulated preheating the oven, to 375° F and then placing the soup in the oven for 5 to 10 minutes, when in the kitchen some 20 minutes were required to achieve a nice top-baked effect. Given our small sample size, we are in no position to assess whether the required preparation times are systematically inaccurate. Generative A.I. fabrications occur in unexpected places, but will have significant impact where precise instructions are expected. Perhaps the design of GPT-4 responds to this problem by omitting some preparation details. Additionally, we did not evaluate the reported nutritional composition information supplied by the A.I. chef.

Overall, however we observed in both GPT-3.5 and GPT-4 a culinary logic across the instruction steps that did lead in fact to tasty meals, even if our final output, not arranged for food photography - its very own generative system - never appeared as appetizing as the DALL-E images.

### 5.2   Thinness of Generative A.I.

From the perspective of design practices, the most important barrier to enthusiasm for generative A.I. is compromised aesthetic quality and a lack of refinement. Overall, the synthetic outputs seem derivative of mediocre examples, suggesting a lack of curatorial ability on the part of GPT generators. This is perhaps most perceivable in the image materials and the contexts they invoke, with high-resolution 'thinness' [38] on display, good enough for stock imagery. Given the rapid pace of change in A.I., with each new version of GPT systems portending to move closer to Artificial General Intelligence - A.I. with human-like abilities - the current thinness might morph into thicker forms of cultural production. Who can know?

Others have also observed, and lamented, that derivative generative artifacts risk normalizing shoddy work. Perhaps not immediately, but the 'standardization of mediocrity' [39] may have corrosive effects that are difficult to predict. As mentioned above, scale matters in machine learning. Scale serves to improve the performance of generative systems. However, scaling in the form of mass distribution of shoddy products accelerates the standardization of mediocrity and negatively impacts the perception of generative A.I.

### 5.3   Questions, Answers, Fabrications

Academic codes of conduct instill the value of novel ideas and respect for the work of other scholars, artists and researchers. Detailed citation of sources is key to acknowledging the prior contributions and contextualizing personal accomplishments. Working with generative systems adds a new dimension to citation culture, and unsettles some of its fundamental tenets. How can one acknowledge the efforts of unknown workers und proprietary sources? We discussed various approaches to this new condition. The term 'collaboration' was found to be one possible descriptor, even though the collaborators, perhaps underpaid workers, remain anonymous and unrewarded. Moreover, we did not find a way to acknowledge the mediated use of proprietary datasets applied in the training of the models we interacted with, nor were we able to assess if the fancy recipes from the collaboration infringed on existing culinary creations.

While large language models store factual knowledge in their parameter space, and can be fine-tuned to specific tasks to improve the representation of facts, they are still, currently, considered inferior to task-specific architectures [40]. Prompt fine-tuning has been proposed as a method by which to address this shortfall. Still, the consequences of generative systems tendency to fabricate, ranging from minor slippages to inconsistencies as we have observed in the cooking instructions, to falsehoods, will impact different application areas in different ways. Some domains, such as Law [41] and health care, are more sensitive to fabrications than others. While the Neural Recipe Box's imagery is superficially seductive and the scope of inventiveness surprising, the pleasurable culinary arts may suffer from inconsistencies produced by generative systems in unexpected ways. Not everything that looks good is good.

Learning how to interact with a generative system, including prompt engineering, is a new skill students must acquire. While our multi-component prompt configuration was quite elaborate, there are a variety of additional prompt manipulation opportunities, including offering a model a bribe for a better solution, or threats of penalization [42], we have yet to explore. Beyond prompt trickery, knowledge making in exchange with a machine requires not just prompt engineering, but reflection on what to ask for in the first place. The ability to formulate a question receives new-found significance. Perhaps this condition will not remain purely instrumental and be recognized for the opportunity it is, namely to reinvigorate the Socratic art of questioning, recognized long before the advent of generative systems as a viable knowledge-seeking method [43].

Updating scholarly practices for co-existence with generative A.I. is very much work in progress, and will likely continue to morph as the abilities of generative systems evolve. Some suggestions, such as not uploading personal information to generative A.I. systems and citing the tools used, including the version of these tools, are sure to take hold. Other ideas are more contested. For example, one scholar suggests students themselves are now responsible for the output of the tools they deploy. Similar to customers required to perform the work of scanning their shopping items at automated checkout counters, students may be asked to police the very systems [44] that are supposed to facilitate and accelerate discovery.

### 5.4    A.I. in Better Places

Culinary culture has attracted the attention of machine learning researchers long before our experiment [45][46]. However, GPT-3+ generative systems allow the combinatorial exploration of kitchen culture with much greater ease.

Even before the most current intensification, some researchers have come to the conclusion that A.I. is best understood as the newest member of general purpose technologies [47], [48], a class of technologies that impact many sectors of the economy. Once a technology enters this stage, it creates strong linkages to other systems and becomes indispensable. Generative A.I. is here to stay.

If we have identified some places where A.I. is perhaps less suitable, if not wrong, then where might we want to have A.I., specifically in the realm of art and design practices? If nothing else, the mediocrity of A.I. digital visual art should be an impetus to consider alternatives. Certainly manual image production, from painting to photography, will be revitalized by the thinness of generative A.I. art. But is there another answer to the A.I. explosion that does not revert to nostalgia?

Instead of using A.I. to imitate existing modes of cultural production, A.I. could perhaps be applied to create new ones. It could create stimulating experiences humans have not yet been able to generate. At least as far as the Neural Recipe Box goes, it seems more interesting to pursue the opportunity of expanding food culture in lieu of imitating derivative food photography. For example, a generative A.I. recipe machine could consider locally sourced, seasonally available produce, check current pricing and availability at markets with additional attention to fair practice produce and labor practices, contact city food waste services for possible high quality leftovers, query food kitchens, and then suggest fabulous meals conditioned on all those interwoven criteria in a beautiful act of logistics magic no human being may be able to perform.

The cornucopia of generative A.I. assets being put into place by cross-disciplinary endeavors such as the weather simulation machinery *EARTH-2* [49], launched by NVIDIA, are redefining the representational arts. A tighter coupling between A.I. driven simulation and the real world seems inevitable. Artists and designers tuned to these developments will have to find new ways to carve out spaces of inquiry, critique and nuance. We look forward to those opportunities.

**Acknowledgments.** This project was supported in part by an Education grant from Google Research. The course *AI in all the wrong places* is supported in part by the Departments of Art, Media Study as well as Computer Science and Engineering at the University at Buffalo.

## Appendix

Here are the complete GPT3.5 and GPT-4 Neural Recipe Box responses for item 3 in Tables 2 and 3, based on the following user selections:

| | |
|---|---|
| Dish Type: Soups | Cuisine Type: French cuisine |
| Food Restrictions: Gluten-free | Meal Type: Dinner |
| Mood: Excited | Texture: Soft |
| Smell: - | Color: Blue |
| Temperature: Freezing (0 °C or lower) | Heating Method: Roasting |
| Calories: 100-200 calories | |

**GPT-3: Lentil Soup**

Ingredients:
- 1 cup of dried lentils
- 2 carrots, peeled and chopped
- 2 stalks of celery, chopped
- 1 onion, chopped
- 3 cloves of garlic, minced
- 1 tablespoon of olive oil
- 4 cups of chicken broth
- 2 bay leaves
- 1 teaspoon of dried thyme
- Salt and pepper to taste

Safety Assurance:
This recipe is safe and edible as it follows the recommended guidelines for cooking and food preparation. Please ensure that you follow the instructions carefully and use fresh ingredients.

Instructions:
1. In a large saucepan, heat the olive oil over medium heat.
2. Sauté the onions, celery, and carrots until the onions become translucent, about 5 minutes.
3. Add the garlic and sauté for another minute.
4. Add the chicken broth, bay leaves, and thyme to the saucepan and bring the mixture to a boil.
5. Once boiling, add the lentils and let the soup simmer until the lentils are tender, about 30 minutes.
6. Remove the bay leaves and season the soup with salt and pepper to taste.
7. Preheat the oven to 375°F.
8. Transfer the soup to a roasting pan and roast for 5-10 minutes until the top is slightly browned.

Origin and Cultural Significance:

Lentils originated in the Middle East and have been used in French cuisine for centuries. In France, lentils are known as "les lentilles" and are a popular ingredient in various dishes, including soups, salads, and stews. These legumes are a significant source of protein and essential nutrients, making them a staple food in many cultures worldwide.

Nutritional Composition:

Micronutrients:
- Vitamin A: 19% of Daily Value (DV)
- Vitamin C: 12% of DV
- Calcium: 4% of DV
- Iron: 11% of DV

Macronutrients:
- Protein: 12g
- Fat: 3g
- Carbohydrates: 34g
- Fiber: 13g
- Sugars: 3g

Nutritional Value and Medical Benefits:

Lentils are rich in several essential nutrients, including protein, fiber, and iron. These legumes have been associated with a lower risk of chronic diseases such as heart disease, diabetes, and certain types of cancer. They also help lower cholesterol levels and improve digestive health.

User Comments:
1. Positive - "This soup is delicious! The lentils add a nice crunch, and the roasting gives it a unique flavor. Definitely a recipe I'll be making again!"
2. Negative - "I followed the instructions exactly, but my soup came out very watery. I wish there were more specific measurements for the broth and vegetables."
3. Mixed - "While the flavor was excellent, I found the texture to be too soft for my liking. Next time, I'll try adding some crunchy croutons on top."

**GPT-4: La Soupe Bleue de Lentilles**

Ingredients:
- 1 cup of dried lentils (french: lentilles)
- 1 quart of vegetable stock
- 1 chopped onion
- 4 cloves of garlic
- 1 teaspoon of vegetable oil
- 1 dollop of gluten-free blue food coloring
- Salt and pepper to taste

Instructions:
For the adventurous and spirited food connoisseur, delve into the gradations of French culinary finesse. Roast the lentils over a medium flame, stirring continuously until crispy. Sauté the onions and garlic in vegetable oil till golden brown, emanating their delightful umami aroma of freshly fried comfort food. Subsequently, this mixture is to be added to the stock with the well-roasted lentils and allowed to simmer till the lentils are wickedly soft. Finally, infuse your creation with a hint of adventurous excitement by incorporating gluten-free blue food colouring. Serve this riveting potage freezing to enjoy its exquisiteness.

Safety Assurance:
This recipe is 100% gluten-free and cooked at healthy temperatures. The ingredients used are nutrient rich and safe for consumption.

Origin and Cultural Significance:
Lentils, or 'lentilles' as they say in French, are an integral part of French culinary heritage, particularly in the inland region of Le Puy, known for its high-quality green lentils. The lentil soup - initially a simple stew of lentils and meat - got a modern twist with this blue hue and cold temperature to literally cool and electrify your palate.

Nutritional Composition:

| | |
|---|---|
| Protein | High |
| Fiber | High |
| Carbohydrates | Moderate |
| Fats | Low |
| Iron | Good amount |
| Vitamin B6 | Decent |
| Magnesium | Adequate |

Nutritional Value and Medical benefits:
   According to Healthline, lentils are packed with beneficial nutrients like fiber, protein, minerals and vitamins, they are also extremely low in fat, and contain a high amount of antioxidants. They are considered heart-healthy as they contain important nutrients that can lower 'bad' LDL cholesterol and increase 'good' HDL cholesterol. Consuming lentils can improve and regulate bowel movements and promote a healthy gut environment due to their high fiber content.

Comments:
   1. Anonymous: I absolutely loved this version of lentil soup. The blue color was exciting and distinct. However, I would have preferred more seasoning.
   2. Maria Roux: This was not my cup of tea. The texture was too soft for me and I was not able to enjoy the soup being served cold.
   3. Jacques Dubois: The recipe was interesting and different, however, I feel it could be better. Maybe reducing the roast of lentils a bit could help. But I admire the creativity and it's low-calorie.